\documentstyle{mn}

\input{epsf}
\begin{document}

\title[
A wide-field K-band survey - II. Galaxy clustering
]
{
A wide-field K-band survey - II. Galaxy clustering
}

\author[C.M. Baugh {\it et al.}]
{C.M. Baugh, J.P. Gardner,  C.S. Frenk \&  R.M. Sharples
 \\
University of Durham, 
Department of Physics, Science Laboratories, South Road, Durham DH1 3LE
}
\maketitle

\def\mpc {h^{-1} {\rm{Mpc}}}
\def\and  {{\it {et al.}}}
\def\rmd {\rm d}
\newcommand{\xibar}{\overline{\xi}}
\newcommand{\beq}{\begin{equation}}
\newcommand{\eeq}{\end{equation}}
\def\spose#1{\hbox to 0pt{#1\hss}}
\def\simlt{\mathrel{\spose{\lower 3pt\hbox{$\mathchar"218$}}
     \raise 2.0pt\hbox{$\mathchar"13C$}}}
\def\simgt{\mathrel{\spose{\lower 3pt\hbox{$\mathchar"218$}}
     \raise 2.0pt\hbox{$\mathchar"13E$}}}

\begin{abstract}
We present the first measurement of the angular correlation function 
in a $K$-selected galaxy survey, from two fields covering almost 10 square 
degrees.
The angular correlation function  at small angles is 
 well described by a $\theta^{-0.8}$ 
power law, as for optically selected samples.
The clustering amplitude is reduced  by a factor of $\sim 4$ between galaxies 
with $K<15$ and a fainter magnitude slice with $15<K<16$.
This allows us to 
place constraints upon the redshift distribution of the galaxies and their 
spatial correlation function.
We find no clear evidence for a change in clustering amplitude when galaxies 
are selected by their observed $B-K$ colours. 
\end{abstract}


\section{Introduction}

The measurement of angular correlations in optically selected samples of galaxies 
has been a useful probe of cosmological models and theories of galaxy formation.
The shape of the angular correlation function, $w(\theta)$, on large 
scales measured in the APM Survey (Maddox {\it et al} 1990, 1996) 
showed that there is more structure in the galaxy distribution on large scales 
than is predicted by the standard Cold Dark Matter model.
The scaling of the 
correlation amplitude with magnitude in faint samples has given some 
indication of the nature of faint blue galaxies ({\it e.g.} Efstathiou {\it et al.} 
1991, Roche {\it et al.} 1993).

It is now possible to survey large areas of sky in the near-infrared $K$ band.
Measurement of the angular correlation function allows us to place constraints 
upon the shape of the redshift distribution of $K$-selected galaxies and on 
their spatial two point correlation function.

\section{Measured Correlations}

\begin{figure}
\begin{picture}(100, 350)
\put(0,0)
{\epsfxsize=8.5truecm \epsfysize=12.truecm 
\epsfbox[20 145 575 701]{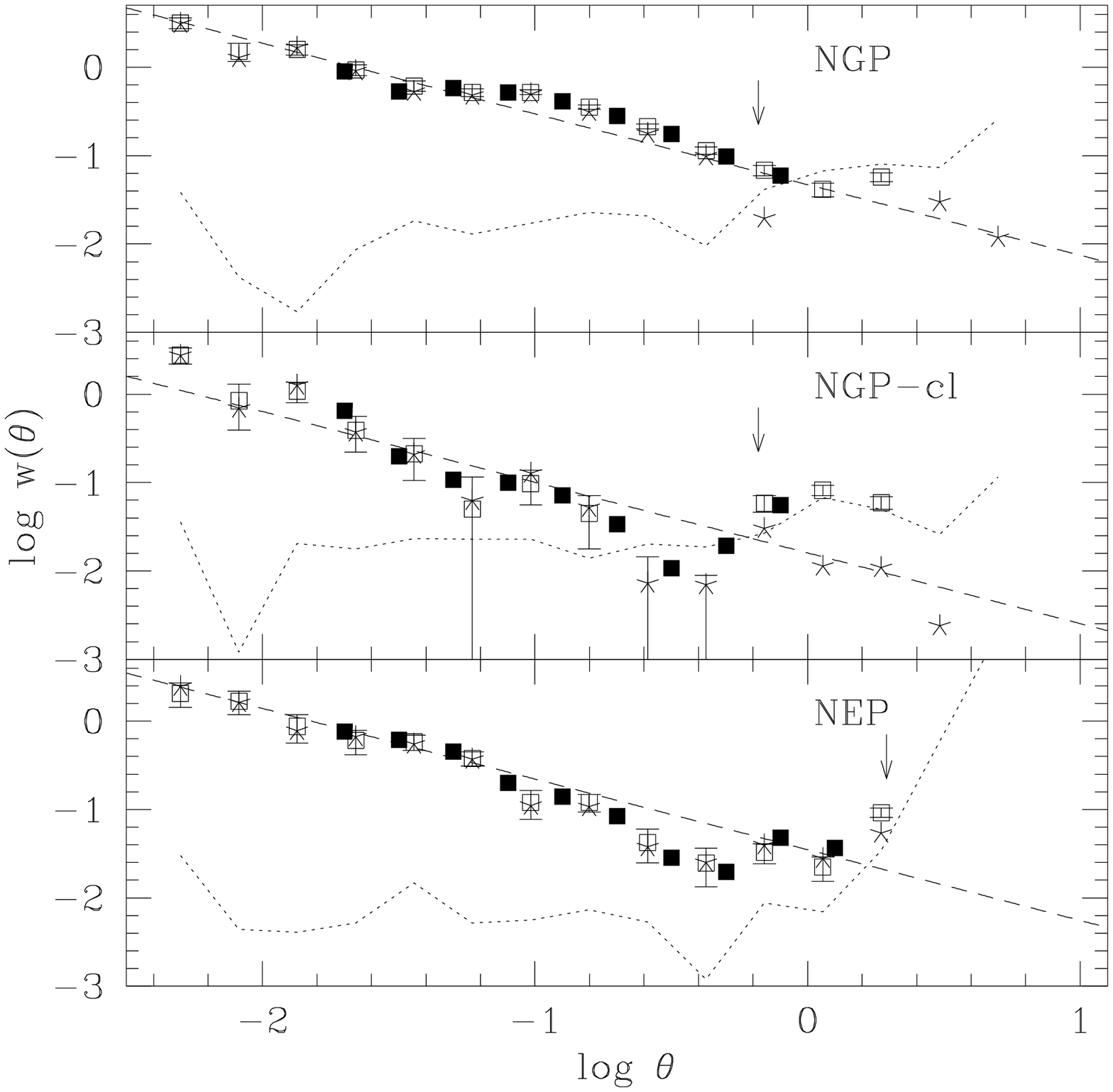}}
\end{picture}
\caption[junk]{
The measured angular correlation functions with the estimated integral 
constraint added for the NEP, NGP and NGP-CL regions for galaxies brighter 
than $K=15$.
The open squares show $w(\theta)$ calculated with the direct estimator, 
the filled squares with the ensemble estimator and the stars with 
Hamilton's estimator.
The error bars on the direct estimator values
are $2 \sigma$ errors calculated as described in the text.
The dashed line shows the power-law fitted to 
the measured $w(\theta)$ on small scales.
The dotted line shows the random-galaxy correlations.
The arrows indicate the smallest linear dimension of each field.
}
\label{fig:k_15}
\end{figure}

\begin{figure}
\begin{picture}(100, 350)
\put(0,0)
{\epsfxsize=8.5truecm \epsfysize=12.truecm 
\epsfbox[20 145 575 701]{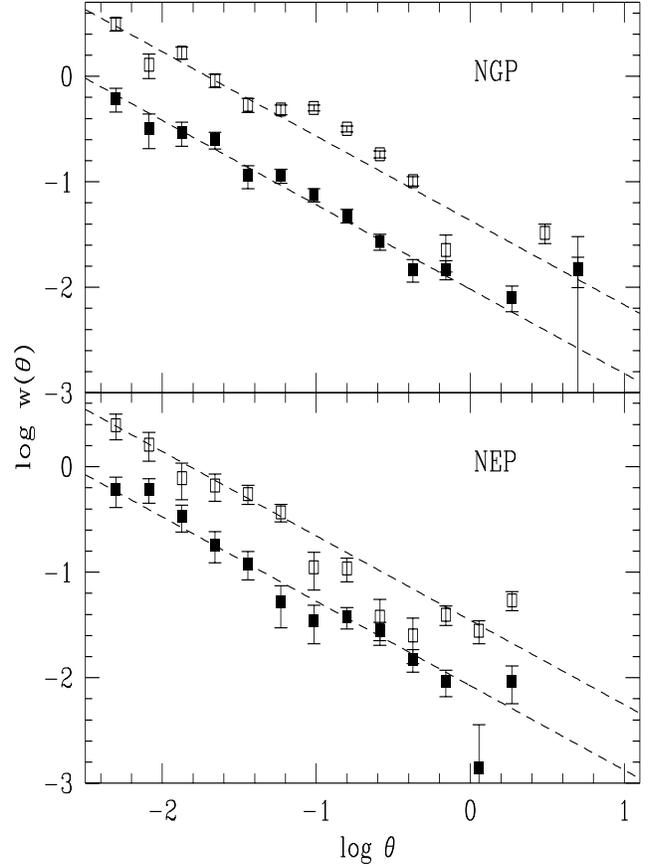}}
\end{picture}
\caption[junk]{
Angular correlation functions for galaxies with $K < 15$ (open 
squares) and $15 < K < 16$ (filled squares), with $2 \sigma$ 
error bars calculated as described in the text.
The dashed lines show the power law $w(\theta)$ fitted at small angles.
The angular correlations are computed using the Hamilton estimator.
}
\label{fig:k_sc}
\end{figure}

\begin{table*}
\begin{center}
\caption[dummy]{Measured angular correlations: The amplitude of a power law 
fit $w(\theta) = A \theta^{-0.8}$ is given along with the $2 \sigma$ range.
The final column gives the value of the integral constraint.}
\label{tab_fit}
\begin{tabular}{ccccccc}
\hline
\hline  
sample    & field  & number of   &  A   &  A (high) & A(low) & $\sigma^{2}$\\
          &        & galaxies    &      &         &        &           \\\hline
$K<15$    & NGP    &  907   &  0.043    & 0.047   & 0.039  & 0.041     \\
          & NGP-CL &  653   &  0.017    & 0.021   & 0.011  & 0.016     \\ 
          & NEP    &  675   &  0.035    & 0.039   & 0.029  & 0.042    \\ \hline
$15<K<16$ & NGP    & 2641   &  0.0096   & 0.0108  & 0.0088 & 0.0093   \\
          & NGP-CL & 2144   &  0.0081   & 0.0093  & 0.0069 & 0.0081    \\ 
          & NEP    & 2122   &  0.0084   & 0.0098  & 0.0070 & 0.0102   \\ \hline
\end{tabular}
\end{center}
\end{table*}

\begin{figure}
\begin{picture}(100, 350)
\put(0,0)
{\epsfxsize=8.5truecm \epsfysize=12.truecm 
\epsfbox[20 145 575 701]{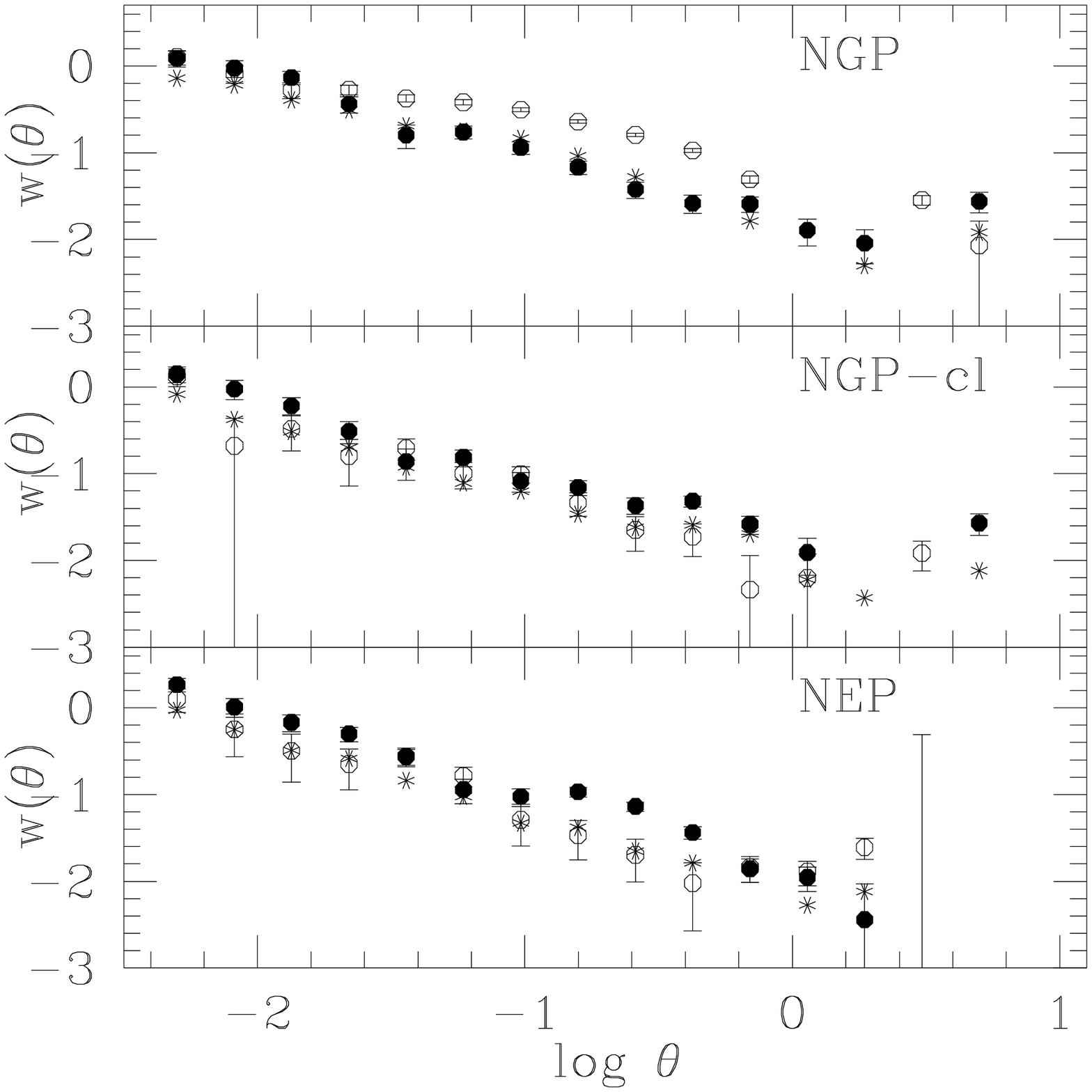}}
\end{picture}
\caption[junk]{
Angular correlation functions for galaxies selected by $B-K$ colour 
from the $K<16$ sample.
The stars show $w(\theta)$ for all galaxies.
The filled circles show $w(\theta)$ for 
galaxies redder than the median $B-K$ colour, 
while open circles show $w(\theta)$ for galaxies 
bluer than the median $B-K$.
The error bars are $2 \sigma$ errors calculated as described in the 
text.
}
\label{fig:k_col}
\end{figure}

We have measured the clustering of galaxies in two fields  imaged 
in the $K$ band with a NICMOS3 detector and in the $B$, $V$ 
and $I$ bands using a $2048^{2}$ CCD camera.
Details of the image reduction can be found in Gardner 
{\it et al} (1996 in preparation).
The star/galaxy separation for objects with $K<15$, $I<18$, 
$V<19$ and $B<20$ was carried out using a combination 
of image morphology and colour selection (Gardner 1995),
and each galaxy identification was confirmed by eye.
In the range $15 < K < 16$, the star/galaxy separation was 
performed using colour alone.
The NEP field measures $1.97^{\circ} \times 2.58^{\circ}$, while 
the NGP field is rectangular with sides $6.91^{\circ} \times 0.66^{\circ}$.
The NGP field contains a galaxy cluster, so we have also estimated 
the angular correlations in this field with the $\sim 0.7$ square degrees 
containing the cluster removed. This effectively produces two 
disjoint regions which we designate NGP-CL.

The angular correlations were measured using three different 
estimators, which we shall refer to as the direct $w_{d}(\theta)$, 
Hamilton $w_{h}(\theta)$ (Hamilton 1993) and ensemble 
$w_{e}(\theta)$ estimators.
In order to compute the direct and Hamilton estimates, 
$20 000$ points were put down at random within the survey regions.
The ensemble estimator is calculated by dividing the fields into 
a grid of cells of side $0.02^{\circ}$.
The form of the estimators is given by 

\begin{eqnarray}
w_{d} (\theta) &=& \frac{2 N_{R}}{(N_{G} - 1)} 
                    \frac{\langle DD \rangle}{\langle DR \rangle} - 1 \\
w_{h} (\theta) &=& \frac{4 \langle DD \rangle \langle RR \rangle}
                   {\langle DR \rangle^{2}} - 1 \\
w_{e} (\theta) &=& \frac{\langle N_{i} N_{j} \rangle}
                 {\langle N_{i} \rangle \langle N_{j} \rangle} - 1, 
\end{eqnarray}
where $N_{G}$ is the number of galaxies in the field that satisfy the 
specified magnitude limits and $N_{R}$ is the number of random points.
The angular brackets indicate an average over all pairs of 
points in the sample for the case of $w_{d}$ and $w_{h}$, 
and an average over pairs of cells  for the case of $w_{e}$,  
that are separated by angle $\theta \pm \delta \theta$. 
The quantities $DD$, $DR$ and $RR$ are the number of distinct 
data-data, data-random and random-random pairs at angular separation  
$\theta$. 
The number of galaxies in cell $i$ is given by $N_{i}$.

The direct estimator suffers from linear biases arising from spurious 
correlations between the survey window and the data and is 
restricted by the accuracy with which the mean density of galaxies is known 
(Hamilton 1993, Landy \& Szalay 1993).
The Hamilton and ensemble estimators do not suffer from such a bias.
The mean galaxy density is determined from the galaxy counts in each field, 
and so all the estimators will be biased low with 
respect to the true angular correlations by an amount 

\begin{equation}
\sigma^{2} \approx \frac{1}{\Omega^{2}} 
\int {\rm d}\Omega_{1} {\rm d}\Omega_{2} w_{T}(\theta_{12}),
\label{eq:ic}
\end{equation}
where $w_{T}(\theta)$ is the true angular correlation function and 
$\Omega$ is the solid angle of the field (Groth \& Peebles 1977).
We estimate the size of this `missing variance' or integral 
constraint by assuming that the true angular correlations are 
given by $w_{T}(\theta) = A \theta^{-0.8}$ and performing a 
Monte-Carlo integration of equation (\ref{eq:ic}) over each field.
The amplitude $A$ is then set by fitting the measured correlation function 
on small angular scales to (Roche {\it et al} 1993): 

\begin{equation}
w(\theta) = A \theta^{-0.8} - \sigma^{2}.
\end{equation} 
We have also estimated the magnitude of the integral constraint by 
calculating the variance in the number of galaxies found in mock versions 
of our fields drawn from a much larger simulated APM catalogue 
(Gazta\~{n}aga \& Baugh 1996).
The integral constraint found in this way is approximately $25 \%$ 
smaller, which indicates the error resulting from the assumption that 
$w(\theta)$ is a power law up to scales corresponding to the angular 
size of the field.

The angular correlations for galaxies with $K<15$ 
are shown in Figure \ref{fig:k_15} for each field. 
Throughout this paper we plot the measured correlations plus the 
estimated value of the integral constraint, given in the final column of 
Table \ref{tab_fit}.
We have measured $w(\theta)$ in 
logarithmic bins of width $\Delta \log \theta = 0.2$.
The dotted line shows the angular correlation between the random points in 
each field and the data which biases the direct estimator.
The dashed line shows the power-law correlation function given by 
a least squares fit to the measured correlations on scales of 
$0.005^{\circ} - 0.06^{\circ}$, as listed in Table \ref{tab_fit} 
for the Hamilton estimator.
The angular correlation function bins are not independent, so we quote 
the $2 \sigma$ range in the fitted amplitude in columns 5 and 6 of 
Table \ref{tab_fit}.
Errorbars are computed using 
$\delta w(\theta)  = \sqrt{(1 + w(\theta))/\langle DD \rangle}$
(Hewitt 1982).
This may underestimate the errors, so we plot the $2 \sigma$ errors 
obtained from this formula.
We have found that the $2 \sigma$ errors computed in this way 
are comparable to the errors obtained with a bootstrap resampling of
the data (Ling, Frenk \& Barrow 1986).

The correlations in the NGP region rise above the power law fitted to 
small scales in the angular range $0.06^{\circ}-0.50^{\circ}$.
The NEP field shows a deficit of pairs on  these scales, compared 
with the fitted power law.
This suggests that this field could be slightly underdense, as a break in 
$w(\theta)$ is not expected at $\theta < 1^{\circ}$  for $K < 15$; in 
the optical, this corresponds to a magnitude limit of approximately 
$B \approx 19$, and $w(\theta)$ is observed to be a power law until 
$\theta \approx 2^{\circ}$ (Maddox, Efstathiou \& Sutherland 1996).
The agreement between the values of $w(\theta)$ measured with the 
different estimators is impressive in the NEP and NGP fields.
There are some discrepancies, however,  
between the estimators in the NGP-CL field 
due to the hole caused by omitting the cluster, which 
results in less reliable measurements of $w(\theta)$ in this field.
From now on we shall plot only the angular correlation function obtained 
from the Hamilton estimator.

The angular correlation function in the $K<15$ sample is compared with that 
in the deeper $15 < K < 16$ sample in Figure \ref{fig:k_sc}, where we have 
plotted $w_{h}(\theta)$, again with $2 \sigma$ errors. 
The dashed lines show the power-law correlation function fitted to the 
measurements at small angles.
There is a reduction in the amplitude of clustering by a factor of $\sim 4$
in the $15 < K < 16$ sample compared with the brighter sample.
The correlations in the deeper NEP and NGP samples are better described 
by a power-law on all scales, suggesting that the sampling 
fluctuations which affect the brighter catalogue are no 
longer significant for $K<16$.

Finally, we examine whether the clustering strength 
depends on galaxy colour.
We find no consistent difference in the amplitude of 
$w(\theta)$ in each field when we split the $K < 16$ catalogue 
into `red' and `blue' galaxies at the median $B-K$ colour. 
Figure \ref{fig:k_col} compares the angular correlation function in the 
blue and red subsamples 
with the correlations for the whole $K < 16$ sample.
In the NGP field, 
 blue galaxies do appear to be more strongly clustered than both the 
red galaxies and the full sample.
However, the angular correlations in this field are dominated 
by the presence of the cluster.
In the region that contains the cluster, there are $\sim 2$ times more 
galaxies with blue colours compared with the number that have red colours.
In the NEP field, the red galaxies appear to be more strongly clustered than 
the blue galaxies, in particular around $\log \theta \sim -0.6$.

\section{Interpretation}

\begin{table}
\begin{center}
\caption[dummy]{The correlation length needed to reproduce 
the amplitude of clustering in the NEP field, 
for a range of theoretical redshift distributions and 
for two different cases of the evolution of clustering with redshift}
\label{tab:modamp}
\begin{tabular}{ccccc}
\hline \hline
redshift     & \multicolumn{2}{c} {$K < 15$} & \multicolumn{2}{c} {$15<K<16$} \\
distribution & $r_{0}$ & $\epsilon$          & $r_{0}$ & $\epsilon$  \\ \hline
optical      &  6.8    &  0.0                & 4.5     & 0.0         \\
fit          &  6.3    & -1.2                & 4.0     & -1.2        \\ \hline
passive      &  7.5    &  0.0                & 6.1     & 0.0         \\
evolution    &  6.9    & -1.2                & 5.2     & -1.2        \\ \hline
no evolution &  6.3    &  0.0                & 4.3     &  0.0        \\
             &  5.8    & -1.2                & 3.8     & -1.2        \\ \hline
\hline  

\end{tabular}
\end{center}
\end{table}

\begin{figure}
\begin{picture}(100, 350)
\put(0,0)
{\epsfxsize=8.5truecm \epsfysize=12.truecm 
\epsfbox[20 145 575 701]{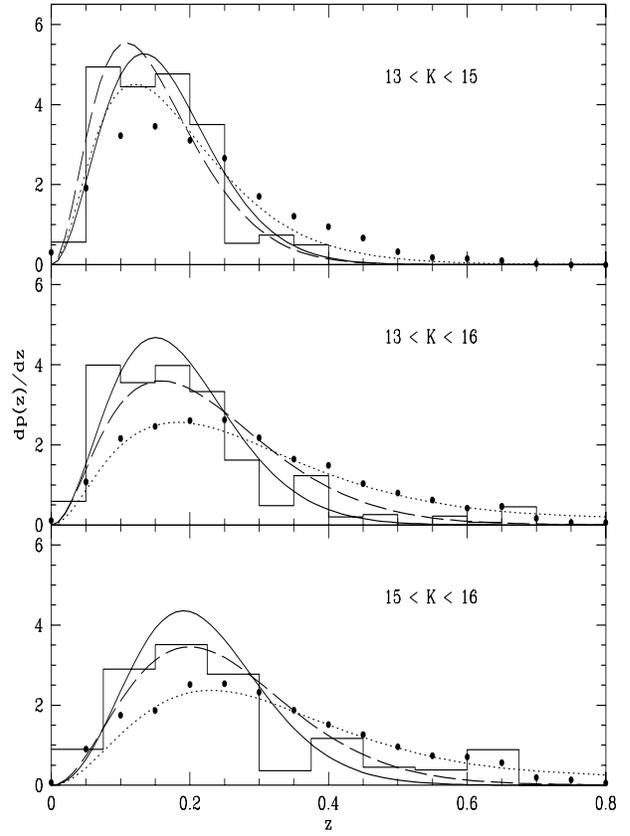}}
\end{picture}
\caption[junk]{
The histograms show the redshift distribution combined from the surveys 
of Songaila {\it et al.} (1994) and Glazebrook {\it et al.} (1995).
The solid curve shows the parametric fit described in the text.
The dashed curve shows the prediction of a no evolution model and the 
dotted curve shows a model with passive evolution.
The circles show the predictions of the semi-analytic model for 
galaxy formation of Cole {\it et al } (1994).
}
\label{fig:dndz}
\end{figure}

The angular correlation function is related to the 
spatial correlation function $\xi(r)$ and the redshift 
distribution of the galaxies by Limber's equation 
(Peebles 1980, \S 56).
On small scales, both $w(\theta)$ and $\xi(r)$ are measured 
to have power law forms ({\it e.g.} 
Davis \& Peebles 1983, Maddox {\it et al} 1990, 1996, 
Baugh 1996) and Limber's equation 
can then be written in the form (Efstathiou {\it et al.} 1991):

\begin{equation}
w(\theta)  = \sqrt{\pi} \frac{\Gamma( (\gamma-1)/2 )}{\Gamma(\gamma/2)}
\frac{A}{\theta^{\gamma - 1}} r_{0}^{\gamma}, 
\label{eq:wtheory}
\end{equation}
where
\begin{equation}
A = \int_{0}^{\infty} g(z) \left( \frac{{\rm d} N}{{\rm d} z} \right)^{2} 
{\rm d} z /
\left[ \int_{0}^{\infty} ( {\rm d} N / {\rm d} z) {\rm d} z \right]^{2} 
\end{equation}
and
\begin{equation}
g(z) = ({\rm d} z /{\rm d} x) x^{1 - \gamma} F(x) (1 + z)^{-(3 + \epsilon 
- \gamma)},
\end{equation}
where $F(x)$ depends upon the cosmology and is equal to unity for $\Omega=1$. 
Here we have parameterised the redshift evolution of the spatial 
correlation function as
\begin{equation}
\xi(r, z) = \left( \frac{r_0}{r} \right)^{\gamma} (1 + z)^{-(3 + \epsilon)}.
\end{equation}
If clustering is fixed in proper coordinates then $\epsilon = 0$.
If the pattern of clustering is fixed instead in comoving coordinates, then 
$\epsilon \approx -1.2$ for $\gamma \approx 1.8$ (Efstathiou {\it et al } 
1991).

The redshift distribution of $K$-selected samples is less 
well determined than that of optical samples.
In order to model the redshift distribution, we have combined 
redshifts from the Hawaii $K$-band Survey (Songaila 
{\it et al} 1994) and from Glazebrook {\it et al.} (1995), weighting each galaxy 
by the field area, sampling rate and redshift completeness fraction as described 
in Baugh \& Efstathiou (1993).
We have used the total $K$ magnitudes listed in Table 3 of Songaila 
{\it et al} and the $20 h^{-1} {\rm kpc}$ metric aperture magnitudes 
from Table A1 of Glazebrook {\it et al.}
Figure \ref{fig:dndz} shows the observed redshift distribution for several 
cuts in $K$ magnitude.
The solid curve shows the prediction of the parametric form adopted by 
Baugh \& Efstathiou (1993) as a fit to the redshift distribution of 
optically selected galaxies 
\begin{equation}
({\rm d} N/{\rm d} z) {\rm d} z 
= \frac{3 {\cal N}(m) \Omega_{s}}{2 z_{c}^{3}}
z^{2} \exp( -(z/z_{c})^{3/2}) {\rm d} z, 
\end{equation}
where the median redshift is $z_{m} = 1.412 z_{c}$.
For the combined $13 < K < 15$ dataset, we find a median redshift of 
$z_{m} \sim 0.155$, whilst for the deeper slice $13<K<16$ we obtain $z_{m} \sim 0.175$.
Also plotted are the predicted redshift distributions for 
`passive' (dotted line) and `no evolution' (dashed line) models constructed with 
the Bruzual and Charlot (1993, 1996 in preparation) 
spectral energy distributions described in Gardner et al (1996) and 
Gardner (1996).
The circles show the predictions of the {\it ab initio} semi-analytic 
model for galaxy formation described by Cole {\it et al} (1994).
The theoretical curves give reasonable fits to the combined dataset, 
except perhaps at higher redshifts, where in any case the observed 
redshift distribution is less well determined.

A further constraint on the form of the redshift distribution is the 
scaling of the amplitude of $w(\theta)$ with apparent magnitude between 
the $K<15$ and $15<K<16$ samples.
In Table \ref{tab:modamp} we find the values of the correlation length 
$r_{0}$ that reproduce the observed amplitude of the angular correlation 
function in the NEP field, for two specific cases of the evolution 
of clustering, $\epsilon=0$ and $\epsilon=-1.2$.
Local optically selected samples give $r_{0} \sim 5 h^{-1} {\rm Mpc}$, 
though the real space spatial correlation function $\xi(r)$ has a shoulder 
feature between $4 < r < 25 h^{-1} {\rm Mpc}$ (Guzzo {\it et al} 1991, 
Baugh 1996, Ratcliffe {\it et al.} 1996), which complicates fitting 
a simple power law to $\xi(r)$.

\section{Summary} 

We have presented angular correlation functions, $w(\theta)$, 
for galaxies selected in the  $K-$band over a large area of sky.
We find that $w(\theta)$ is well fitted by a $\theta^{-0.8}$ power law on 
small angular scales, as observed for optically selected galaxies.
There is a factor of $\approx 4$ reduction in the amplitude of clustering
for a faint magnitude slice, $15<K<16$, compared with 
galaxies with $K<15$.
This behaviour can be reproduced assuming the redshift 
distribution predicted by a simple 
no evolution model, if the correlation length of galaxies with a median 
redshift of $z \sim 0.23$ is around $4.3 h^{-1} {\rm Mpc}$ 
whilst for galaxies at 
a lower median redshift of $z \sim 0.13$, $r_{0} \sim 6.3 h^{-1} {\rm Mpc}$, 
and clustering is fixed in proper coordinates.
We find no consistent change in the strength 
of clustering in the different fields 
when galaxies are selected by 
their observed $B-K$ colours, but this is does not rule out a dependence of 
clustering amplitude on colour in samples of higher median redshift (Landy {\it et al.} 1996).

The measured clustering contributes to the 
uncertainty in the number counts (Gardner {\it et al} 
1996) because different patches of sky contain different numbers of galaxies depending 
upon whether they are in high or low density regions.
The integral constraint $\sigma^{2}$ gives an indication of the fractional variance 
in the number counts (see Peebles 1980 \S36).
Taking into account the fact that we have two widely separated fields, the rms variance 
at $K<15$ is $\sigma_{rms} = 0.14$ whilst for $15<K<16$ $\sigma_{rms} = 0.07$.
These values are upper limits, though tests with simulated catalogues that 
have the same clustering show that these numbers would be reduced by $\sim 0.02$.
The contribution to the variance in the counts from galaxy clustering is 
several times larger than that expected from Poisson statistics alone.

{\small 

\section*{Acknowledgements}
We would like to thank Esperanza Carrasco for assisting with the 
data collection. We acknowledge useful conversations with Richard Fong.
This work was supported by a PPARC rolling grant 
for Extragalactic Astronomy and Cosmology at Durham.

\setlength{\parindent}{0mm}

\bigskip

{\bf REFERENCES} 
\bigskip

\def\refe {\par \hangindent=.7cm \hangafter=1 \noindent}
\def\aj { ApJ, }
\def\aa {A \& A, }
\def\ajs{ ApJS, }
\def\mn { MNRAS, }
\def\apl { ApJ, }

\refe Baugh, C.M., Efstathiou, G., 1993, \mn 265, 145
\refe Baugh, C.M., 1996, \mn 280, 267
\refe Bruzual, G.A., Charlot, S., 1993, \aj 405, 538
\refe Bruzual, G.A., Charlot, S., 1996, in preparation
\refe Cole, S., Aragon-Salamanca, A, Frenk, C.S., Navarro, J.F., Zepf, S.E., 
      1994, \mn 271, 781
\refe Davis, M., Peebles, P.J.E., 1983, \aj 267, 465
\refe Efstathiou, G., Berstein, G., Katz, N., Tyson, J., Guhathakurta, P., 1991, \aj 380, L47
\refe Gardner, J.P., 1995, \ajs 98, 441
\refe Gardner, J.P., 1996, \mn 279, 1157
\refe Gardner, J.P., Sharples, R.M., Carrasco, B.E., Frenk, C.S., 1996 \mn submitted
\refe Gazta\~{n}aga, E., Baugh, C.M., 1996 in preparation
\refe Glazebrook, K., Peacock, J.A., Miller, L., Collins, C.A., 1995, 
      \mn 275, 169
\refe Groth, E.J., Peebles, P.J.E., 1977, \aj 217, 38
\refe Guzzo, L., Iovino, A., Chicarini, G., Giovanelli, R., Haynes, M.P., 1991, \aj 
      382, L5
\refe Hamilton, A.J.S., 1993, \aj 417, 19
\refe Hewitt, P.C., 1982, \mn 201, 867
\refe Landy, S.D., Szalay, A.S., 1993, \aj 412, 64
\refe Landy, S.D., Szalay, A.S., Koo, D.C., 1996 \aj 460, 94
\refe Ling, E.N., Frenk, C.S., Barrow, J.D., 1986, \mn 223, 21p
\refe Maddox, S.J., Efstathiou, G., Sutherland, W.J., Loveday, J., 1990, 
      \mn 242, 43p
\refe Maddox, S.J., Efstathiou, G., Sutherland, W.J., 1996, \mn submitted
\refe Peebles, P.J.E., 1980 Large Scale Structure of the Universe, Princeton
\refe Ratcliffe, A., Shanks, T., Broadbent, A., Parker, Q.A., Watson, F.G., 
      Oates, A.P., Fong, R., Collins, C.A., 1996, \mn submitted	
\refe Roche, N., Shanks, T., Metcalfe, N., Fong, R., 1993, \mn 263, 360
\refe Songaila, A., Cowie, L.L., Hu, E.M., Gardner, J.P., 1994, \ajs 94, 461

}
\end{document}